\documentclass[conference,table,xcdraw,10pt,final]{IEEEtran}

\usepackage[utf8]{inputenc}

\usepackage[T1]{fontenc}
\usepackage{flushend}
\usepackage{amsmath}
\usepackage{bm}

\PassOptionsToPackage{hyphens}{url}
\usepackage{hyperref}

\usepackage[hyphenbreaks]{breakurl}

\usepackage[detect-weight=true, detect-family=true,per-mode=symbol-or-fraction,binary-units=true]{siunitx}
\DeclareSIUnit\sample{Sa}
\DeclareSIUnit\cycle{cycle}
\DeclareSIUnit\instruction{instruction}
\DeclareSIUnit\core{core}

\usepackage{subfig}

\usepackage[table,xcdraw]{xcolor}

\usepackage{dblfloatfix}

\usepackage{tabularx}

\usepackage{multirow}

\usepackage[export]{adjustbox}

\usepackage[absolute]{textpos}
\textblockorigin{1.7cm}{26cm}

\title{Energy Efficiency Aspects of the\\AMD Zen 2 Architecture}

\author{\IEEEauthorblockN{Robert Sch\"{o}ne\textsuperscript{1}
	\hfill
Thomas Ilsche\textsuperscript{2}
	\hfill
Mario Bielert\textsuperscript{2}
	\hfill
Markus Velten\textsuperscript{2}
	\hfill
Markus Schmidl\textsuperscript{3}
	\hfill
Daniel Hackenberg\textsuperscript{2}}
	\IEEEauthorblockA{
		\textit{Center for Information Services and High Performance Computing (ZIH)}, TU Dresden, Dresden, 01062, Germany\\
\textsuperscript{1} robert.schoene@tu-dresden.de \hfill
\textsuperscript{2} firstname.lastname@tu-dresden.de  \hfill
\textsuperscript{3} markus.schmidl@mailbox.tu-dresden.de
		}
}

	\newcommand{\mycite}[1]{\textit{``#1''}}
	
	\ifx\DraftModeOn\undefined
	
	\newcommand{\todoti}[1]{}
	\newcommand{\todomb}[1]{}
	\newcommand{\todors}[1]{}
	\newcommand{\tododh}[1]{}
	\newcommand{\todomv}[1]{}
	\newcommand{\todoms}[1]{}

	\newcommand{\figref}[1]{Figure~\ref{fig:#1}}
	\newcommand{\tabref}[1]{Table~\ref{tab:#1}}
	\newcommand{\secref}[1]{Section~\ref{sec:#1}}

	\newcommand{\papertarget}[2]{}

	\else
	
	\newcommand{\todoti}[1]{\todo[color=yellow!60,inline,size=\small]{Thomas: #1}}
	\newcommand{\todomb}[1]{\todo[color=cyan!60,inline,size=\small]{Mario: #1}}
	\newcommand{\todors}[1]{\todo[color=green!60,inline,size=\small]{Robert: #1}}
	\newcommand{\tododh}[1]{\todo[color=red!60,inline,size=\small]{Daniel: #1}}
	\newcommand{\todomv}[1]{\todo[color=orange!60,inline,size=\small]{MarkusV: #1}}
	\newcommand{\todoms}[1]{\todo[color=blue!60,inline,size=\small]{MarkusS: #1}}
	
	\newcommand{\figref}[1]{\textcolor{red}{Figure~\ref{fig:#1}}}
	\newcommand{\tabref}[1]{\textcolor{red}{Table~\ref{tab:#1}}}
	\newcommand{\secref}[1]{\textcolor{red}{Section~\ref{sec:#1}}}

	\newcommand{\papertarget}[2]{\todo[inline]{Target: #1: \\Deadline: #2}}
	
	\fi

\begin{document}

	\maketitle
	
	\begin{abstract}
		In High Performance Computing, systems are evaluated based on their computational throughput.
		However, performance in con\-tem\-po\-rary server processors is primarily limited by power and thermal constraints.
		Ensuring operation within a given power envelope requires a wide range of sophisticated control mechanisms.
		While some of these are handled transparently by hardware control loops, others are controlled by the operating system.
		A lack of publicly disclosed implementation details further complicates this topic.
		However, understanding these mechanisms is a prerequisite for any effort to exploit the full computing capability and to minimize the energy consumption of today's server systems.
		This paper highlights the various energy efficiency aspects of the AMD Zen 2 microarchitecture to facilitate system understanding and op\-ti\-miza\-tion.
		Key findings include qualitative and quantitative descriptions regarding core frequency transition delays, workload-based fre\-quen\-cy limitations, effects of I/O die P-states on memory performance as well as discussion on the built-in power monitoring capabilities and its limitations.
		Moreover, we present specifics and caveats of idle states, wakeup times as well as the impact of idling and inactive hardware threads and cores on the performance of active resources such as other cores.

	\end{abstract}
	
	\begin{figure}[bp]
		\centering
		\includegraphics[width=0.99\columnwidth]{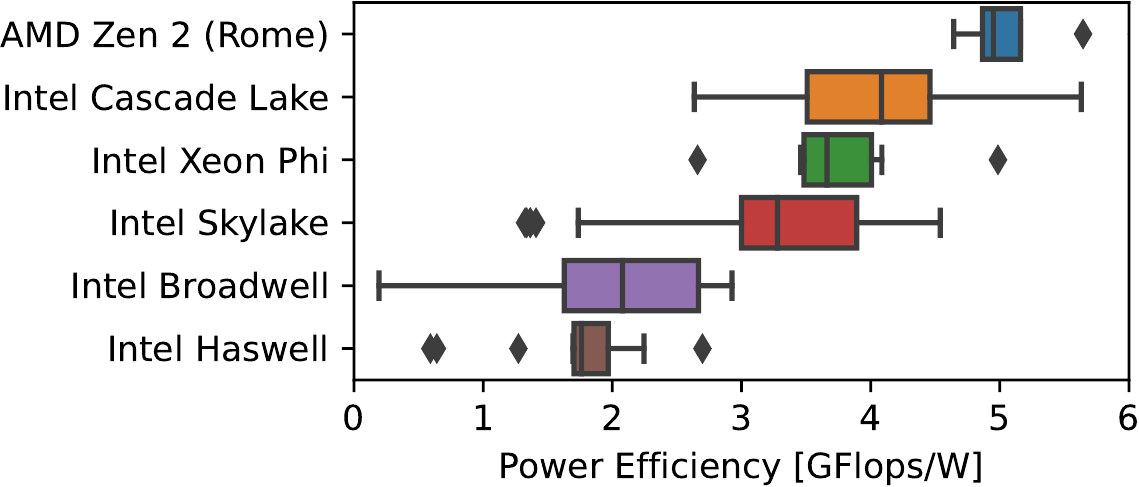}
		\caption{\label{fig:green500}Efficiency of systems with x86 processors in the 2021/07 Green500 list~\cite{TOP500} (architectures with $> 5$ systems).}
	\end{figure}%

\begin{IEEEkeywords}
AMD; Zen 2; Epyc Rome; power saving; energy efficiency; DVFS; C-State; performance; RAPL
\end{IEEEkeywords}

\begin{textblock}{1}(0,0)
\textbf{© 2021 IEEE under \href{https://doi.org/10.1109/Cluster48925.2021.00087}{DOI 10.1109/Cluster48925.2021.00087}.} Personal use of this material is permitted. Permission from IEEE must be obtained for all other uses, in any current or future media, including reprinting/republishing this material for advertising or promotional purposes, creating new collective works, for resale or redistribution to servers or lists, or reuse of any copyrighted
component of this work in other works.
\end{textblock}

	\section{Introduction}
	
	\label{sec:intro}

	With the Epyc Rome processor generation, AMD processors gained a noticeable share in the TOP500 list of supercomputers for the first time since Opteron Interlagos, which debuted in 2011.
	The new architecture is not only competitive in terms of performance, but also power efficiency among systems using general-purpose x86 processors as \figref{green500} shows.

	The Green500~\cite{TOP500} list, used to create \figref{green500}, ranks top High Performance Com\-puting (HPC) systems by their energy efficiency under full load.
	In practice, \emph{energy efficiency} is not only defined by power ef\-fi\-cien\-cy during peak performance.
	Many mechanisms, such as P-states or C-states, concern operation during scaled-down performance or idle phases.
	Other mechanisms, such as Turbo frequencies and power capping, aim at maximizing performance under thermal and power constraints.
	They are supported by internal energy measurements, which can also be used for energy-efficiency optimizations.
	This paper analyzes these dynamic and highly configurable mechanisms rather than the application-specific \emph{performance per watt}.
	The re\-sult\-ing insight is the foundation to improve the complex interactions between applications, operating systems (OSs), and independent hardware control for performance and energy efficiency.
	
	The paper is structured as follows:
	\secref{rel} and \secref{arch} discuss existing work on the evaluation of energy efficiency mechanisms and the Rome architecture, respectively.
	\secref{test-system} introduces our test system and power measurement infrastructure.
	The next three sections highlight particular aspects, each including methodologies, test results, and a discussion:
	In \secref{freq}, we unveil details on processor frequencies.
	\secref{idle} covers characterizations of processor idle states.
	In \secref{rapl}, we validate the accuracy of the internal power monitoring mechanism.
	We conclude the paper with a sum\-ma\-ry and outlook in \secref{summary}.

	\section{Related Work on Efficiency Mechanisms}
	\label{sec:rel}

	\begin{figure*}[b]
		\centering
		~\hfill
		\subfloat[\label{fig:arch-ccd}Core Complex Die (CCD) with Core Complexes (CCXs)]{
			{\includegraphics[width=.75\columnwidth]{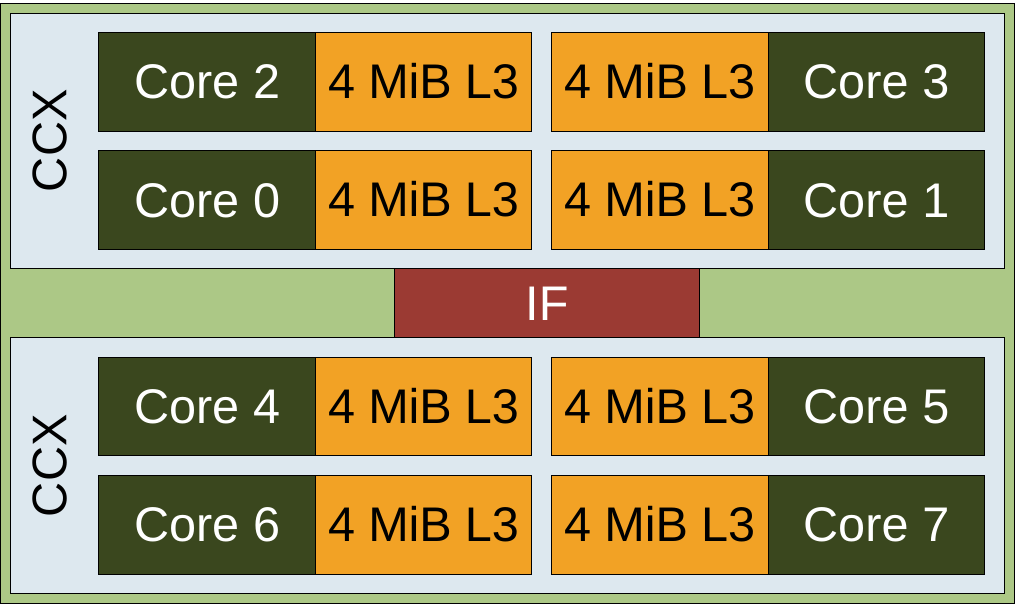} %
			}
		}
		\hfill~\hfill
		\subfloat[\label{fig:arch-io-ccd}I/O die with memory controllers (UMC), attached memory, IF-Switches (brown), CCDs, repeaters, and I/O; xGMI attachements not depicted; based on~\cite{AMD_PPR_Server}]{
			{\includegraphics[width=0.9\columnwidth]{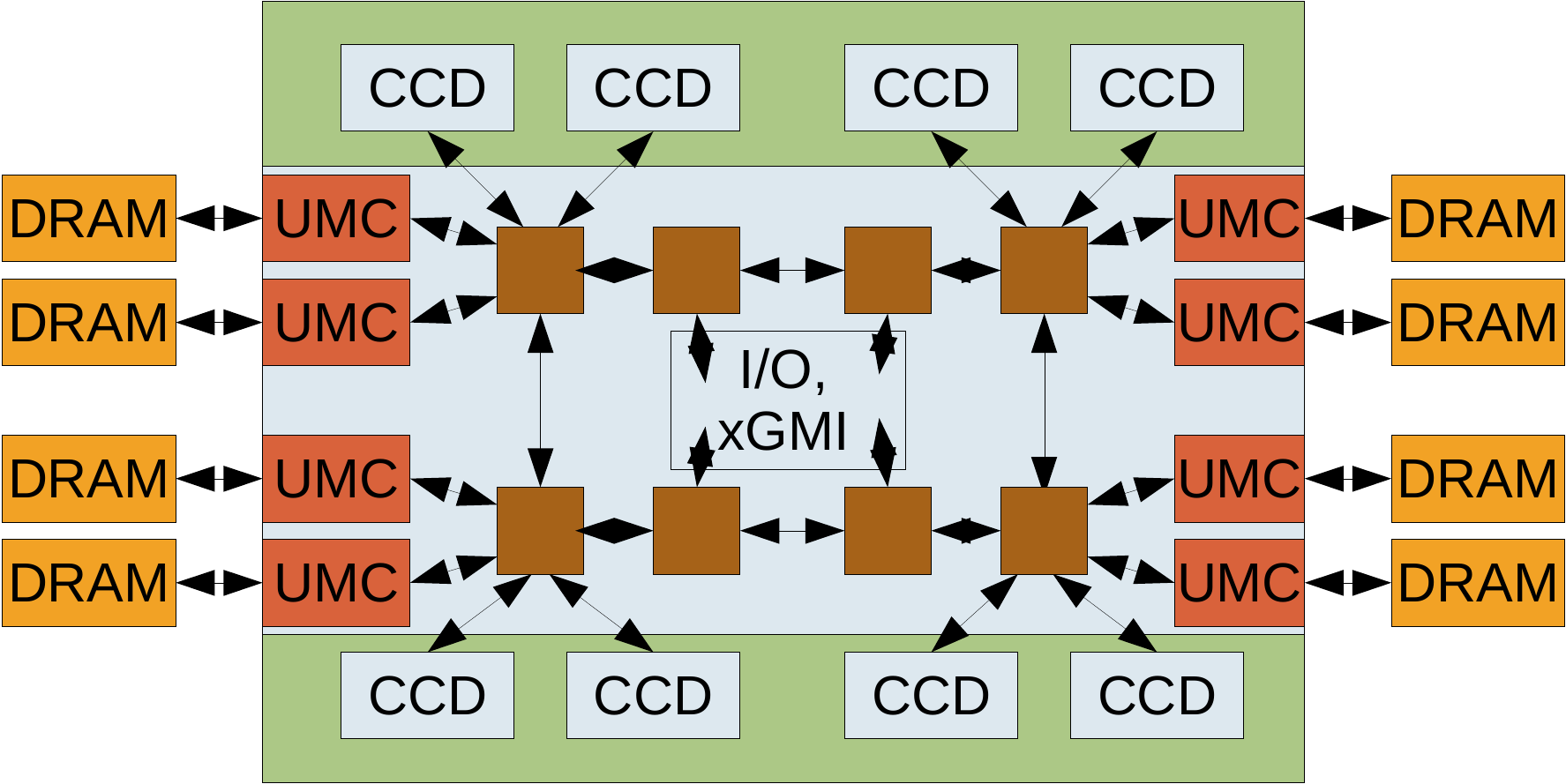}
			}
		}
		\hfill~
		\caption{\label{fig:arch}Block diagram of AMD Rome Architecture, communication via Infinity Fabric (IF)}
	\end{figure*}
	\subsection{ACPI States}
	Power saving interfaces are defined in the Advanced Con\-fig\-u\-ra\-tion and Power Interface (ACPI)~\cite{acpi}.
	Performance states (P-states) provide different performance levels and can be selected during runtime~\cite[Section 2.6]{acpi}.
	Usually processors implement these with Dynamic Voltage and Frequency Scaling (DVFS).
	However, their particular implementation is highly processor-architecture dependent.
	Mazouz et al. were one of the first to investigate this in~\cite{Mazouz_PState}:
	They describe how P-state transition times depend on initial and target processor frequency.
	We show in~\cite{Schoene_2014_CStates} that waking an idling processor core is also frequency-dependent, but additionally depends on the waker-wakee-relation and the applied idle state.
	Likewise, we also show in~\cite{Ilsche_2018_Cstate} how long it takes to enter an idle state.
	Finally, we describe the effect of clock modulation (throttling) in~\cite{Schoene_2016_Tstates}.
	This paper covers P-state and C-state transitions in \secref{freq} and \secref{idle}, respectively.
	Software controlled clock modulation, however, is not publicly documented for the Zen 2 architecture~\cite{AMD_PPR_Server, AMD_PPR_Consumer}.
	
	\subsection{Processor-internal Power Measurement and Capping}
	Processor-internal power monitoring is used to select turbo frequencies and implement power capping~\cite{Rountree_2012_PowerConstraint,Schuchart_2016_PowerVariations}.
	The accuracy of these monitoring mechanisms therefore directly influences processor performance.
	We describe the Intel Running Average Power Limit  (RAPL) for Intel Sandy Bridge and the AMD Application Power Management (APM) for AMD Bulldozer in~\cite{Hackenberg_Power_2013}.
	We find that both are based on models that use data from processor internal resource usage monitors.
	We also analyze RAPL for Intel Haswell processors~\cite{Hackenberg_2015_Haswell} and describe it to be accurate and based on measurements.
	H{\"a}hnel et al. measure the update rate of RAPL in~\cite{Haehnel_2012_RAPL} as \SI{1}{\milli \second}.
	Lipp et al. show in~\cite{Lipp_2020_Platypus} that RAPL can provide significantly higher rates for the core power domain (pp0) for certain processors.
	They leverage this for a side channel attack~\cite{Lipp_2020_Platypus}.
	
	\subsection{Processor-specific Overviews}
	We describe the Intel Haswell server architecture in detail~\cite{Hackenberg_2015_Haswell}.
	In addition to the previously mentioned RAPL analysis, the authors found an asynchronous mechanism that sets core frequencies in an interval of \SI{500}{\micro\second}.
	We also describe the interaction between core and uncore frequency mechanisms and show how concurrency and frequencies influence memory bandwidths.
	Gough et al. provide a broad overview of the Haswell architecture as well as suggestions for tuning server systems according to user requirements~\cite{Gough_2015_EEServers}.
	We describe the Intel Skylake server architecture~\cite{Schoene_2019_SKL}, covering the internal hardware control for uncore frequencies, AVX-frequency mechanisms, and the influence of data on the power consumption of a well-defined workload.
	We show that uncore frequency changes can occur every \SI{1.5}{\milli\second}.

	\section{Architectural Details of ``Rome'' Processors}
	\label{sec:arch}

	\subsection{General Architecture Details}
	Zen~2 uses a modular design on multiple levels~\cite[Section 1.8.1]{AMD_PPR_Server}.
	The structure of the processor is depicted in~\figref{arch}.
	Four cores are clustered in one Core Complex (CCX, also CPU Complex).
	One Core Complex Die (CCD) comprises two CCXs.
	Up to eight CCDs are attached to one I/O die on processors with up to \num{64} cores.
	Based on the core count, two or one of the CCDs attach to the same switch within the I/O die network.
	Each of the switches on the I/O die that connects the CCDs also attaches a memory controller with two memory channels, which can result in four non-uniform memory access (NUMA) nodes.

	Each core has a common front-end which fetches instructions for two independent hardware threads~\cite{AMD_Zen2_Overview}.
	The fetch window is \SI{32}{\byte} wide and fed to a 4-way decoder.
	The back-end is split into two parts:
	One part comprises four Arithmetic Logical Units and three Address Generation Units (AGUs), the other contains two \num{256}-bit-wide Floating-point Multiply-Add (FMA) and two \num{256}-bit-wide floating-point add units.
	The AGUs can be used for two loads and one store per cycle, where each of these can transfer up to \SI{32}{\byte} of data.
	
	Each processor core holds an op cache for \num{4096} ops, \SI{32}{\kibi\byte} L1I and L1D caches, and \SI{512}{\kibi\byte} L2 cache, which is used for instructions and data.
	In addition, each CCX holds \SI{16}{\mebi\byte} of L3 cache, distributed over four slices with \SI{4}{\mebi\byte} each.

	\subsection{Energy Efficiency Details for ACPI States}
	\label{sec:ee-details}
	The AMD Zen\,2 architecture implements a wide range of power saving mechanisms.
	According to AMD's \textit{Processor Programming Reference}~\cite[Section 2.1.14.3]{AMD_PPR_Server}, a maximum of eight P-states can be defined.
	However, on most systems, the number of available P-states will be lower.
	The actual number can be retrieved by polling the \textbf{P-state current limit} Model Specific Register (MSR).
	The definition of single P-states includes specifications for frequency, the \mycite{expected maximum current dissipation of a single core}, and a \mycite{voltage ID}.
	The latter is not publicly documented.
	A processor core frequency can be higher than nominal when using \textit{Core Performance Boost}.
	No implementation details are disclosed for server architectures.
	For desktop processors, AMD describes Precision Boost, where the frequency can be increased in \SI{25}{\mega\hertz} steps as part of the SenseMI technology\footnote{\url{https://community.amd.com/t5/blogs/understanding-precision-boost-2-in-amd-sensemi-technology/ba-p/416073}}.
	This would match the frequency multiplier entry in the MSR, where multiples of \SI{25}{\mega\hertz} can be defined.
	
	Zen\,2 implements the usage of idle power states with the in\-struc\-tions \texttt{monitor/mwait} and I/O addresses, which when accessed trigger the entering of one of these states.
	The latter are defined in the \textbf{C-state base address} MSR~\cite[Section 2.1.14.3]{AMD_PPR_Server} .
	There is no indication that AMD implements clock modulation as Intel does~\cite{Schoene_2016_Tstates}.
	However, according to Singh et al., processors support more frequency ranges on some market segments with
	run-time duty-cycle settings~\cite[Fig. 2.1.5]{AMD_Core_details}.

	\subsection{Other Energy Efficiency Details}
	In addition to traditional power saving mechanisms, AMD also implements I/O die P-states.
	According to~\cite[Section “ROME” SOC]{AMD_Zen2_Overview}, the frequency is decoupled from core P-states and can be used to control the performance and power budget of the I/O die.
	The reference document~\cite[Section 2.1.14.3]{AMD_PPR_Server} also indicates that the L3 cache has a dedicated frequency domain and names some restrictions (\mycite{L3 frequencies below \num{400}\,MHz are not supported by the architecture}).
	However, the underlying mechanism is not disclosed.
	
	Parts of the core can be clock-gated at a fine granularity even during active states.
	Singh et al. state in~\cite{AMD_Core_details} that \mycite{clock gating opportunities were identified for low-IPC patterns, where only a portion of the pipeline is used}.
	Suggs et al. name \mycite{continuous clock and data gating improvements} in~\cite[Section Energy Efficiency]{AMD_Zen2_Overview}.
	In particular, the upper \SI{128}{\bit} of the SIMD-capable Floating-Point (FP) units were target for optimization, since only specialized software uses \num{256}-bit SIMD instructions.
	Singh et al. state that \mycite{Zen 2 gated the FP clock mesh \num{128}-bit regions with no additional clocking overhead [...]}, which \mycite{saved 15\% clock mesh power in idle and average application cases where FP was inactive}.
	Due to the large number of partially clock-gated, wide super scalar execution units, power consumption now depends on the actual workload being executed.
	\mycite{[A]n intelligent EDC manager which monitors activity [...] and throttles execution only when necessary} helps to avoid peaks that \mycite{cause electrical design current (EDC) specifications to be exceeded}~\cite[Section Floating-Point/Vector Execute]{AMD_Zen2_Overview}.
	We evaluate this in \secref{firestarter}.
	
	Burd et al. describe more power saving mechanisms for AMD Zen/Zeppelin processors in~\cite{Burd_Zeppelin}, which could also be available for AMD Zen 2/Rome.
	These include a package C-state PC6 \mycite{in which the CPU power plane can be brought to a low voltage when there are no active CPU cores}, but also a low power state that could be implemented in the I/O die.
	In this state, \mycite{most of the IO and memory interfaces are disabled and placed in a low-power state}.
	Burd et al. also mention the possibility to lower the infinity fabric link width between sockets.
	
	With the Zen architecture, AMD replaced APM (Application Power Management) with RAPL (Running Average Power Limit)~\cite{AMD_PPR_Server}.
	The implementation seems similar to the Intel solution, but uses different MSRs.
	While Intel typically provides multiple domains and the option to limit power consumption over various time frames~\cite[Section 14.10]{Intel_2020_Manual3}, AMD only describes registers for reading package and core power consumption.
	However, the latter is available with a per-core spatial resolution, compared to per-package for Intels core domain (pp0).
	While Intel switched from a model to measurement with the Haswell architecture,
	slides from AMD indicate that they use a model based on \mycite{> \num{1300} critical path monitors, \num{48} on-die high speed power supply monitors, \num{20} thermal diodes, [and] \num{9} high speed droop detectors}\footnote{Michael Clark, The ``Zen'' Architecture, \url{https://www.slideshare.net/pertonas/amd-ryzen-cpu-zen-cores-architecture}} for Zen desktop processors.
	We evaluate the resulting accuracy of RAPL in \secref{rapl}.
	
	In~\cite{Burd_Zeppelin}, Burd et al. describe how on the Zen architecture, System Management Units (SMUs) work together to communicate applied frequencies and necessary voltages, where each die of the package implements its own SMU.
	In~\cite[Fig. 7]{Burd_Zeppelin}, they describe that from this set of SMUs, a Master SMU is chosen, which evaluates data from other SMUs and runs control loops for package power and temperature.
	It also triggers frequency changes and controls the external voltage regulator.
	The slide set of~\cite{Naffziger_ISSCC}\footnote{%
\url{https://www.slideshare.net/AMD/amd-chiplet-architecture-for-highperformance-server-and-desktop-products}} also shows SMUs being part of the Rome architecture and still responsible for \mycite{power management} and \mycite{thermal control}.

	\section{Test System and Power Measurements}
	\label{sec:test-system}
	For our analysis, we use a dual socket system with two AMD EPYC 7502 processors, where each processor hosts 32 Cores in 4 CCDs.
	We configured the system to use the \mycite{2-Channel Interleaving (per Quadrant)} mode~\cite{AMD_NUMA}.
	From the available frequencies (\SIlist{1.5;2.2;2.5}{\giga\hertz}), we use the reference frequency (\SI{2.5}{\giga\hertz}) and the ``Auto'' I/O die P-state unless specified otherwise.
	By default, memory is clocked at \SI{1.6}{\giga\hertz}.
	The system runs Ubuntu Linux 18.04 with kernel 5.4.0-47-generic.
	We use the GNU Compiler Collection (GCC) in version 7.5.0 as the default compiler.
	Access to MSRs is performed via the \textit{msr} kernel module, except for RAPL energy readouts, for which we use custom libraries\footnote{\url{https://github.com/tud-zih-energy/x86_energy} with \texttt{x86\_adapt} backend}.
	We use the Linux cpufreq governor ``userspace'' to control processor frequencies.
	By default, we enabled all available C-states.
	We use \texttt{sysfs} files to control  C-states\footnote{\texttt{/sys/devices/system/cpu/cpu\textbackslash d+/cpuidle/state[012]}} and hardware threads\footnote{\texttt{/sys/devices/system/cpu/cpu\textbackslash d+/online}}.

	We use a ZES LMG670 power analyzer with L60-CH-A1 channels to measure the total AC power consumption of the test system.
	In our configuration, the power measurement has an accuracy of $\pm$(\SI{0.015}{\percent} + \SI{0.0625}{\watt}).
	During the experiments, a separate system collects the active power values at \SI{20}{\sample\per\second}.
	The out-of-band data collection avoids any perturbation.
	Measurement data is merged with the internal power and performance monitoring in a post-mortem step.
	For quantitative comparisons, we use average power values within the inner \SI{8}{\second} of a \SI{10}{\second} interval in which one workload configuration is executed continuously.
	This approach avoids in\-ac\-cu\-ra\-cies due to misaligned timestamps.
	We pre-heat the system for power-sensitive workloads.
	
	\section{Processor Frequencies}
	\label{sec:freq}
	
	\subsection{Influence of Idling Hardware Threads on Core Frequencies}
	To investigate the influence of idling hardware threads, we set up a simple workload, where one thread of a core executes a constant workload (\texttt{while(1);}) running at minimum frequency (\SI{1.5}{\giga\hertz}).
	We monitor its frequency with \texttt{perf stat -e cycles -I 1000}.
	Then, we change the frequency of the second thread of the same core to the nominal frequency (\SI{2.5}{\giga\hertz}).
	We let the second thread idle and monitor its activity also with \texttt{perf stat}.
	The idling thread reports only a usage of less than \SI{60000}{\cycle\per\second} and uses idling states.
	However, even though the second thread is idling, the frequency of the first thread is elevated to the nominal frequency (\SI{2.5}{\giga\hertz}) rather than its configured minimal frequency (\SI{1.5}{\giga\hertz}).
	In another attempt, we disable the idling thread.
	Still, the frequency of the core is defined by the offline thread.
	Based on this observation, it can be advantageous to set the frequency of unused hardware threads to the minimal frequency to allow active threads to control their effective frequency.
    We never observed this behavior on Intel processors with enabled deep idle states.
	It may therefore be unexpected for system administrators.
	
	\subsection{Frequency Transition Delays}
	While operating systems change processor frequencies based on resource usage, researchers also use these mechanisms to optimize energy efficiency for code paths~\cite{Rountree_2009_Adagio,Vysocky_2020_DVFSopt}.
	However, the possible time scales for both highly depend on the delay of the frequency transition, which can take tens to hundreds of microseconds~\cite{Mazouz_PState,Hackenberg_2015_Haswell}.
	
	For our tests, we refined the approach from~\cite{Mazouz_PState} to measure this delay as follows:
	The benchmark switches the core frequency and measures the runtime of a minimal workload.
	It repeats the measurement until the expected performance of the target frequency is reached.
	Afterwards, the performance is measured another \num{100} times and validated with a confidence interval of \SI{95}{\percent}.
	Then, the benchmark applies the initial frequency, waits for the appropriate performance level and validates it as before.
	If either one of the two validations fails, the sample and the following sample is discarded.
	Before the next measurement, the benchmark waits a random time between \SI{0}{\milli\second} and \SI{10}{\milli\second}.
	We measure each combination of initial and target frequency \num{100000} times.
	Other cores in the system are set to the minimum frequency of \SI{1.5}{\giga\hertz}.

	\begin{figure}[tb]
		\centering
		\includegraphics[width=.8\columnwidth]{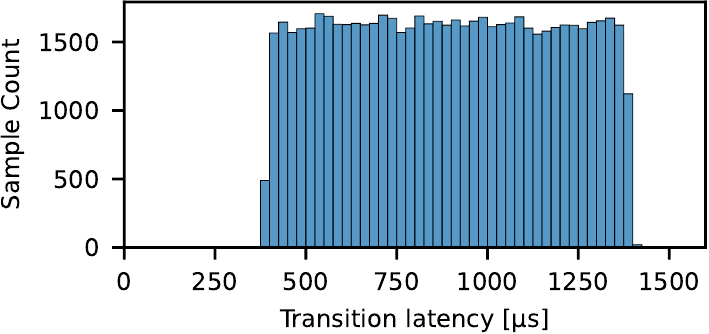}
		\caption{\label{fig:switch-time-random}Histogram of frequency transition delays from \num{2.2} to \SI{1.5}{\giga\hertz} (random starting time, \SI{25}{\micro\second} bins).}
	\end{figure}
	
	\figref{switch-time-random} shows the dis\-tri\-bu\-tion of transition delays for a switch from \SIrange{2.1}{1.5}{\giga\hertz}.
	The measured latencies are approximately uniformly distributed between \SI{390}{\micro\second} and \SI{1390}{\micro\second}.
	This wide dis\-tri\-bu\-tion indicates that an internal fixed update interval of \SI{1}{\milli\second} is used.
	A similar mechanism was observed for core and uncore frequencies of Intel processors~\cite{Hackenberg_2015_Haswell,Schoene_2019_SKL}.
	The delay from the initial request to a transition slot is up to \SI{1}{\milli\second} and the actual frequency change takes another \SI{390}{\micro\second}.

	We also observe comparable results for other frequency com\-bi\-na\-tions with the exception of changes between \SI{2.5}{\giga\hertz} and \SI{2.2}{\giga\hertz}, where we experienced a significantly higher rate of invalid measurements.
	When switching from \SIrange{2.5}{2.2}{\giga\hertz}, some measured latencies are below the assumed minimal transition delay, down to \SI{160}{\micro\second}.
	Symmetrically, there are less measurement samples above \SI{1100}{\micro\second}.
	When switching from \SIrange{2.2}{2.5}{\giga\hertz}, some transitions are executed instantaneously (\SI{1}{\micro\second} delay).
	Here, the previous transition did not finish completely (e.g., frequency set, but not voltage).
	Therefore, returning to a previous setting is faster.
	The effect disappears with random wait times of at least \SI{5}{\milli\second}.

	Based on the measurements, we can show that AMD introduced update intervals for core frequencies that define times when fre\-quen\-cy transitions can be initiated.
	On our system, the period of that window is \SI{1}{\milli\second}, compared to \SI{500}{\micro\second} on Intel systems~\cite{Hackenberg_2015_Haswell,Schoene_2019_SKL}.
	The delay of approx. \SI{390}{\micro\second} for the actual transition (\SI{360}{\micro\second} for increasing frequency) is also significantly higher compared to the Intel Haswell architecture (\SIrange{21}{24}{\micro\second}).
	This might be caused by communication between the SMUs, which likely creates higher delays compared to a centralized Power Control Unit on Intel architectures.

	\subsection{Influence of Mixed Frequencies on a Single CCX}
	\label{sec:freq-mix}
	For this evaluation, we configure cores of a single CCX to use different frequencies.
	We run a simple workload (\texttt{while(1);}) on all cores of a CCX and measure the frequency of one core, which is configured differently than other cores.
	We monitor each setup for \SI{120}{\second} and capture the frequency every second via \texttt{perf stat}.
	The results are presented in \tabref{freqs-mixed}.
	Evidently, core frequencies are reduced if other cores on the same CCX apply higher frequencies.
	While this effect is moderate for a core running on \SI{1.5}{\giga\hertz}, with a reduction of \SI{33}{\mega\hertz} and \SI{71}{\mega\hertz}, the performance penalty for \SI{2.2}{\giga\hertz} is more severe with a reduction of \SI{200}{\mega\hertz}.
	
\begin{figure}[t]
		\centering
		\includegraphics[width=.9\columnwidth]{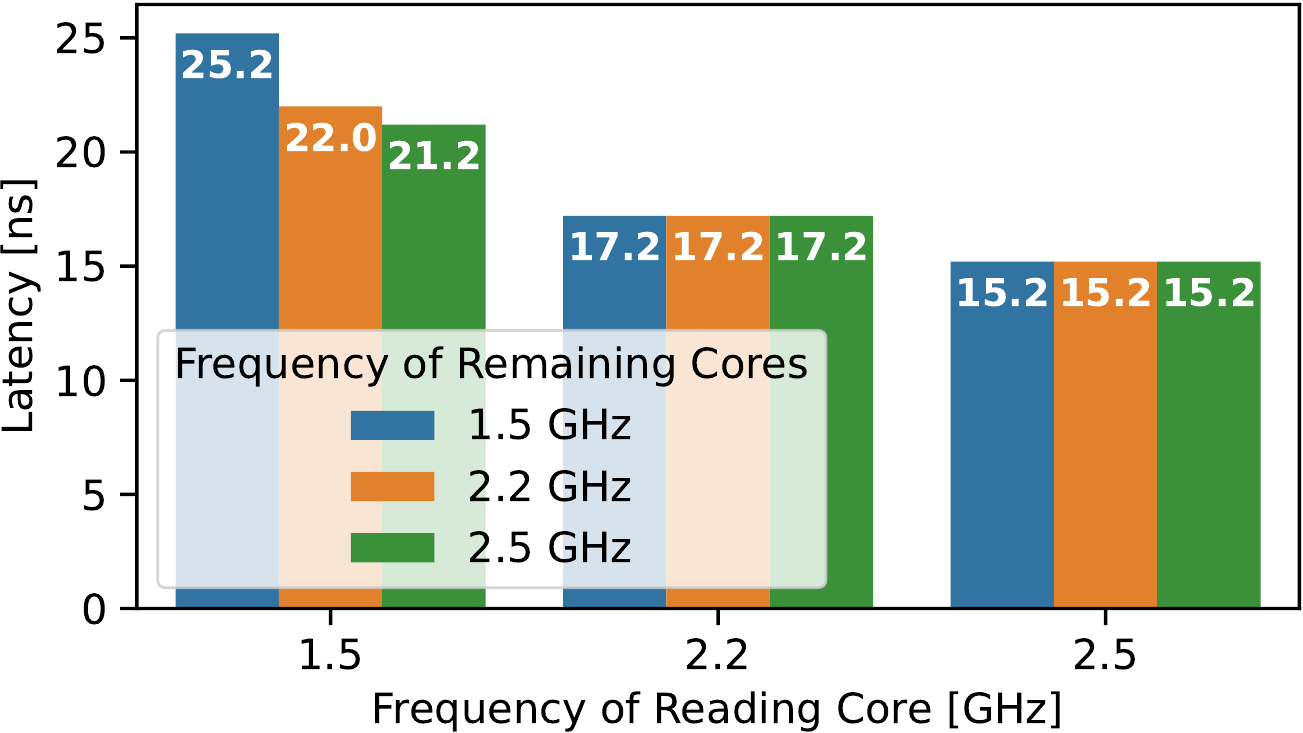}
		\caption{\label{fig:l3-latency}L3-cache latencies in a mixed frequency set-up on one CCX.} %
	\end{figure}

	To understand the influence different core frequencies have on the L3-cache frequency, we use a pointer chasing benchmark, developed by Molka et al.~\cite{x86_membench}.
	We disabled hardware prefetchers in this test and explicitly used huge pages via the \textit{hugetlbfs}.
	As with the previous test, we test one core of one CCX, while the other cores are in an active state.
	We measure each combination several times and present the minimal measured latencies to filter out outliers, where the measurement has been influenced by software (OS) or hardware (processor internal mechanisms).
	As \figref{l3-latency} shows, the latency to the L3 cache decreases for a core running at \SI{1.5}{\giga\hertz}, when other cores in the same CCX apply a higher core frequency, even though its own frequency is decreased by the previously mentioned effect.
	We explain this with an increased L3-cache frequency that is defined by the highest clocked core in the CCX.

	Both effects have severe consequences for performance modeling and energy efficiency optimizations.
	Even if an optimal core fre\-quen\-cy is predicted correctly and applied to a processor core, other cores can disturb the well-optimized setup, resulting in a loss of performance and energy efficiency.
	However, the same mechanism that can reduce the frequency and subsequently performance of a single core in a CCX can also decrease L3-cache latencies and therefore increase performance.

	\subsection{Influence of I/O Die P-state and DRAM Frequency on Memory Performance}
	
 \label{sec:io-pstate}
	In addition to core and L3-cache frequencies, the I/O die has its own voltage and frequency domain, which can influence NUMA, I/O, and memory accesses that pass the I/O die.
	In this section, we analyze how different configurations for I/O die and memory frequencies will influence main memory performance.
	To do so, we use two benchmarks: The STREAM-Triad benchmark proposed by McCalpin~\cite{McCalpin_STREAM}, and the memory latency benchmark described by Molka et al.~\cite{x86_membench}.
	In contrast to previous measurements, we use the Intel compiler for the STREAM benchmark as it reaches higher performance levels.
	We further vary the number of cores that concurrently access memory by using additional well placed threads, defined via OpenMP environment variables.
	We disabled hardware prefetchers, as mentioned in \secref{freq-mix}, and explicitly use huge pages for the latency benchmark.
	For both benchmarks we vary I/O die P-states and DRAM frequencies in the BIOS.

	\begin{table}[t]
		\centering
		\caption{\label{tab:freqs-mixed}Applied mean core frequencies \si{\giga\hertz} in a mixed frequency set-up on one CCX, Tests with lower applied frequency than other cores are highlighted.}
		\begin{tabular}{cc|ccc}
			&                          & \multicolumn{3}{c}{Set frequencies of other cores [\si{\giga\hertz}]}  \\
			&                          & 1.5   & 2.2                     & 2.5                     \\\hline
			\multirow{3}{2cm}{Set frequency of measured core [\si{\giga\hertz}]} & 1.5 & 1.499 & \cellcolor{red!15}{\textbf{1.466}} & \cellcolor{red!15}{\textbf{1.428}} \\
			& 2.2 & 2.200 & 2.199                   & \cellcolor{red!15}{\textbf{2.000}} \\
			& 2.5 & \hspace{.3cm} 2.497 \hspace{.3cm} & \hspace{.3cm} 2.499  \hspace{.3cm}                 & \hspace{.3cm} 2.499     \hspace{.3cm}
		\end{tabular}
	\end{table}
	\begin{figure*}[t]
		\centering
		\subfloat[\label{fig:dram-bw-triad}STREAM Triad bandwidth]{
			\includegraphics[width=.55\textwidth]{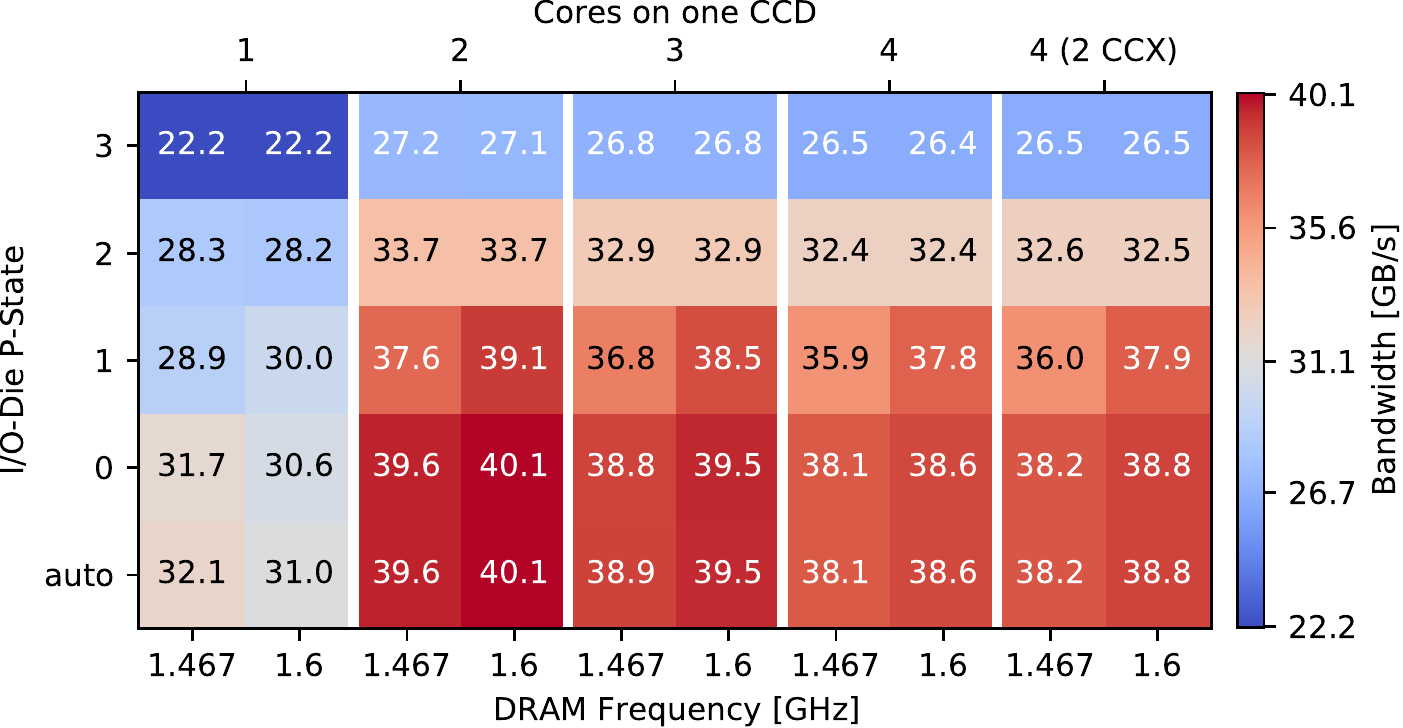}}
        \hfil
		\subfloat[\label{fig:dram-latency}latency]{
			\includegraphics[width=.214\textwidth]{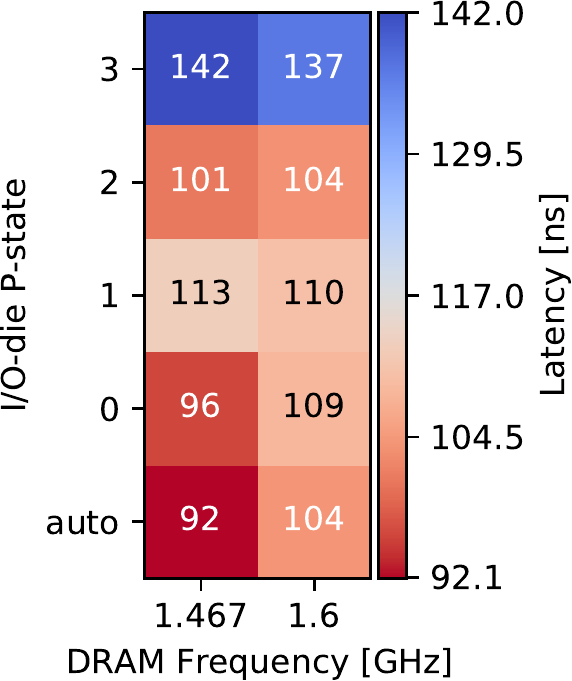}}
		\caption{\label{fig:dram-performance}DRAM bandwidth and latency for I/O die P-states and DRAM frequencies.} %
	\end{figure*}
	
	\figref{dram-performance} presents the results.
	The pattern for memory bandwidths shows that two cores on one CCX already reach the maximal main memory bandwidth and additional cores can lead to performance degradation.
	Using higher I/O die P-states reduces power consumption but also lowers memory bandwidth.
	Surprisingly, a higher DRAM frequency does not increase memory bandwidth significantly.
	As expected, the lowest power state performs best in this benchmark and the auto setting has the same performance.
	This could lead to the assumption that for performance analysis it would be best to pin the I/O die P-state to 0 to remove one source of inpredictability.
	However, when looking at the latencies-result, one can see that auto outperforms the P-state 0 with \SI{92.0}{\nano\second} vs \SI{96.0}{\nano\second}.
	Moreover, for the higher memory frequency, also the I/O die P-state 2 performs better than P-state 0.
	This could be attributed to a better match between the frequency domains for memory and I/O die.
	According to our observations, the auto setting performs good for all scenarios. However, we did not investigate the hardware control loop and how fast it reacts to different access patterns.

	\subsection{Frequency Limitations for High-Throughput Workloads}
	\label{sec:firestarter}
	Starting with the Haswell-EP processor generation, Intel defined AVX frequencies, where workloads that use wide SIMD instructions would use a lower ``nominal'' frequency ~\cite{Schuchart_2016_PowerVariations}.
	\cite[Section Floating-Point/Vector Execute]{AMD_Zen2_Overview} describes such a static assignment a \mycite{simplistic approach to mitigating this issue} and describes that Zen\,2 uses \mycite{an intelligent EDC manager which monitors activity [...] and throttles execution only when necessary}.
	To evaluate how the test system is influenced by workloads that utilize all processor resources to the highest extend, we use FIRESTARTER 2~\cite{Schoene_2021_FIRESTARTER2}.
	The workload schedules up to two \num{256}-bit FMA instructions per cycle accompanied by \num{256}-bit vector loads and stores to different levels of the memory hierarchy.
	To maximize back-end utilization, these instructions are interleaved with integer and logical instructions.
	To utilize the front-end, we increase the size of the inner loop such that it does not fit into the L0 op cache but in the L1I cache.
	This limits the maximal throughput of each core to four instructions per cycle.
	Before we run our tests, we execute FIRESTARTER for \SI{15}{\minute} in order to create a stable temperature.
	We run our tests at nominal frequency for two minutes and measure frequency and throughput with \texttt{perf stat}.
	For power measurements, we use the external AC measurements described in \secref{test-system} as well as RAPL package energy counters.
	We exclude data for the first \SI{5}{\second} and last \SI{2}{\second} to avoid including the initialization phase in addition to clock synchronization issues.
	Performance data for all used threads is collected in \SI{1}{\second} intervals.
	
	\begin{figure}[b]
		\centering
		\includegraphics[width=.9\columnwidth]{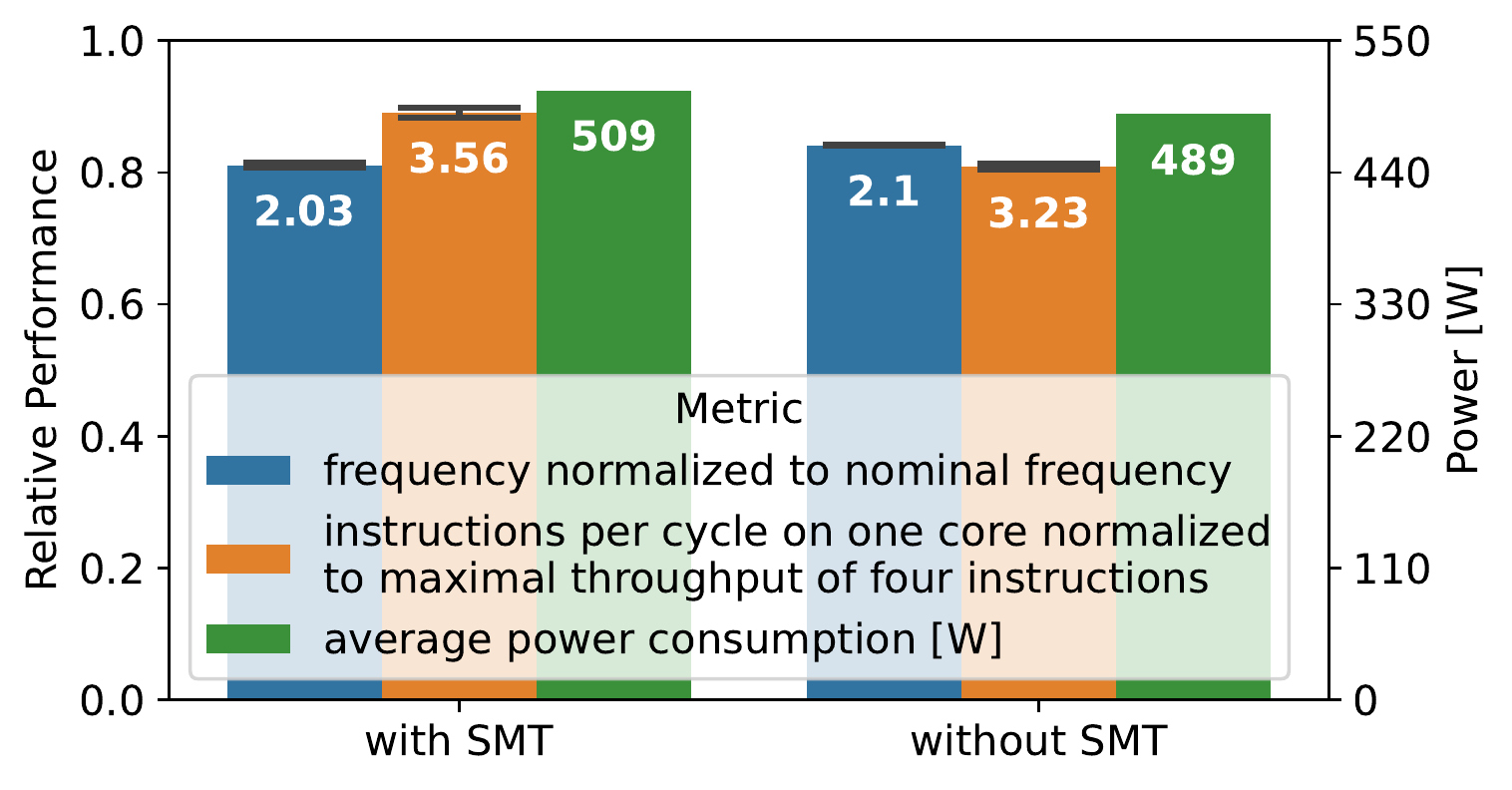}
		\caption{Observed parameters at set nominal frequency \SI{2.5}{\giga\hertz}) with and without the usage of two hardware threads per core. Error bars depict standard deviation.}
		\label{fig:firestarter_frequency}
	\end{figure}
	
	In our measurements, average processor core frequencies are reduced to \SI{2.0}{\giga\hertz}  or \SI{2.1}{\giga\hertz}, depending on whether two threads per core are used or not.
	This is depicted in \figref{firestarter_frequency}.
	The standard deviation is \SI{3.04}{\mega\hertz} and \SI{0.82}{\mega\hertz}, respectively.
	The frequency difference can be explained with a higher average throughput of \SI[per-mode=fraction]{3.56}{\instruction\per\core\per\cycle} (standard de\-vi\-a\-tion \num{0.008}) instead of \SI[per-mode=fraction]{3.23}{\instruction\per\core\per\cycle} (standard deviation \num{0.004}), re\-spec\-tive\-ly.
	Also the average power consumption of the system is higher when both hardware threads of a core are used.
	It is \SI{509}{\watt} instead of \SI{489}{\watt} with a standard deviation of less than \SI{4}{\watt} in both cases.
	Meanwhile, the RAPL package counter reports \SI{170}{\watt} for both processors even though their TDP is stated to be \SI{180}{\watt}.
	This might be related to the accuracy of the processor internal power measurement, which we investigate in \secref{rapl}.
	Enabling Core Performance Boost has almost no influence on throughput, frequency and power consumption.

	Based on our measurements, we conclude that the EDC manager works as expected and will lower processor frequencies if needed.
	This poses a threat to the efficiency of HPC systems.
	In well balanced applications, one throttling processor can slow down the whole program.
	On Intel systems, administrators and users know about the severity of this problem and can apply counter measures (e.g., running highly parallel programs at reduced frequency) based on the documented AVX frequency ranges.
	For AMD Rome systems, measurements are required to determine the actual frequency ranges on a specific processor.

	\section{Power State Details} \label{sec:idle}
	In this section, we analyze processor power states (C-states), as described in~\cite[Section 8.6]{acpi}.
	Idling C-states are triggered by the operating system when there is no thread that can be scheduled on a hardware thread.
	When all threads of a processor core enter such a state, (parts of) the processor core can be clock gated~\cite[Section 5.2.1.1]{Weste_2011} or even power gated~\cite[Section 5.3.2]{Weste_2011}.
	When all cores of a processor are in an idle power state, the processor can take additional measures, e.g., clock and power gating shared components, which lowers the power envelop of the processor further.
	However, returning from an idling power state takes some time, which can violate the requirements for real-time systems.
	To support the decision of the OS, which C-state to use, processors typically hand over ACPI objects describing the transition latency and the average power consumption~\cite[Section 8.4.2.1]{acpi}.
	On our test system, three C-states are supported, the active C-state C0, and two additional idling C-states, C1 and C2\footnote{This paper uses the OS C-state numbering.}.
	While the former is entered with the instructions \texttt{monitor} and \texttt{mwait}, the latter uses IO address 0x814 in the C-state address range described in \secref{ee-details}.
	Transition times are reported as \SI{1}{\micro\second} and \SI{400}{\micro\second}, respectively.
	The power values reported by the hardware to the OS are \texttt{UINT\_MAX} for the active C-state and \texttt{0} for the idle states and cannot contribute towards an informed selection of C-states.

	\subsection{Power Consumption in Different Power States}
	\label{sec:idle-power}

	In the following, we characterize the average AC power consumption of the full system in different configuration of idle states, each measured for \SI{10}{\second}.
	All configurations are shown in \figref{cstate_power}.
	When all hardware threads are using the C2 state to the extent that is possible on a standard Linux system with regular interrupts, the system consumes \SI{99.1}{\watt}.
	In the following experiment, we put more hardware threads in C1 by disabling the C2 state in \texttt{sysfs}.
	The change is applied linearly, following the logical CPU numbering in steps of single CPUs.
	We start with the hardware thread of each core within the first processor package, the second processor package, and then the second hardware threads of each core, again grouped by package.
	With a single core using C1 rather than C2, the power consumption increases by \SI{81.2}{\watt} to a total of \SI{180.3}{\watt}.
	Additional cores in C1 only increase power consumption by \SI{0.09}{\watt} each with no further change when the second hardware thread of each core is put into C1.
	The minor frequency-independent additional power per core is consistent with the observation that the hardware counters for \texttt{cycles}, \texttt{aperf}, and \texttt{mperf} do not advance on cores that are in C1.
	Both effects indicate that cores are clock-gated during C1.
	
	For the active state (C0) we pin an unrolled loop of \texttt{pause} instructions to each hardware thread.
	This workload exhibits a more stable and slightly lower power consumption than \texttt{POLL}, which is also based on \texttt{pause}, but without unrolling and more sophisticated checks for each iteration.
	With one active thread and all others in C2, the system uses almost the same power (\SI{180.4}{\watt}) than with one thread in C1 and all others in C2.
	While C1 and C2 power was independent of core frequency, active power does depend on frequency as expected.
	For \SI{2.5}{\giga\hertz}, additional active cores increase power by \SI{0.33}{\watt} each and hardware thread costs \SI{0.05}{\watt} each.
	On our dual-socket system, there was no measurable impact of activating the second package.
	There appears to be only one criterion for deep package sleep states: All threads of all packages must be in the deepest sleep state.
	The C1 state is only relevant for one specific core, as opposed to the C1E state on Intel systems.

	The reported numbers are only valid for our specific system and depend on the  processor model, processor variations, and other components.
	However, this example particularly highlights the disproportionately high cost of not using the deepest sleep states on a single hardware thread and thus the importance of managing C-states correctly on idle systems.

	Compared with a dual socket Intel system using Xeon Gold 6154 CPUs~\cite[Section III]{Schoene_2019_SKL}, the deepest idle state (\SI{69}{\watt}, all C6) and first core in C1E (\SI[retain-explicit-plus]{+97}{\watt}) are in a similar order of magnitude.
	However, on the Intel Skylake system, each additional active core (\texttt{pause} loop) costs \SI{3.5}{\watt} - about ten times the power of our AMD Rome system.

	\begin{figure}[t]
		\includegraphics[width=\columnwidth]{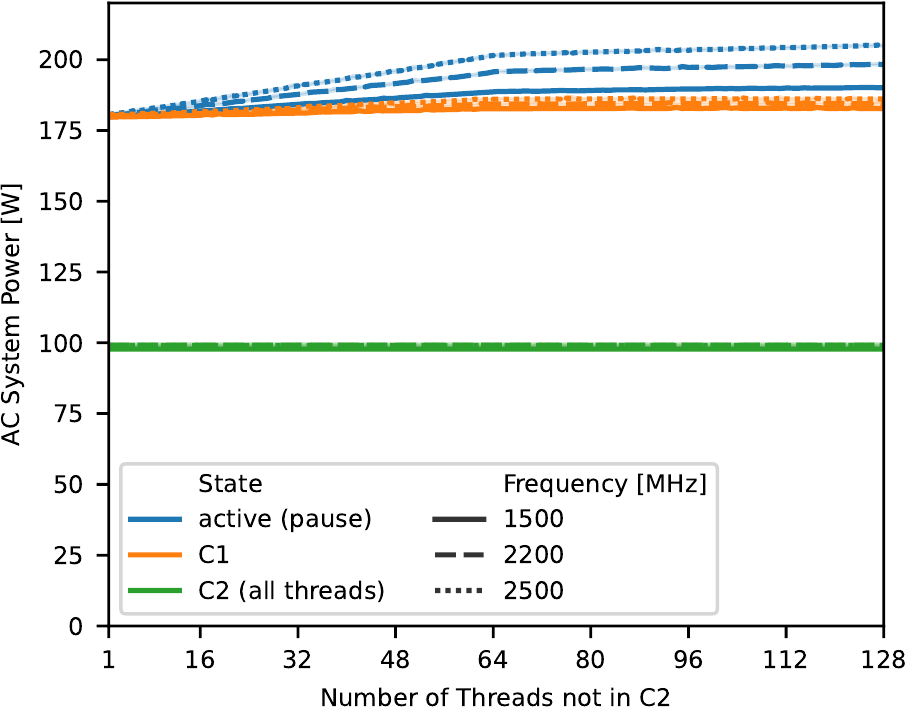}
		\caption{Average full system AC power consumption for different idle combinations with increasing number of hardware threads in lower C-states.}
		\label{fig:cstate_power}
	\end{figure}
	\subsection{Influence of Idling Hardware Threads on Idle States}
	In some scenarios, administrators disable the additional hardware threads of each core via operating system interfaces, to decrease the probability of leaving a package C-state and subsequently increasing the average power consumption during idle times.
	While such an optimization can be recommended on Intel systems, we would strongly discourage using this option on AMD Rome.
	Under conditions we could not yet clearly identify, a strange behavior was observable: even though C2 states are active and used by the active hardware threads, system power consumption is increased to the C1 level as long as the disabled hardware threads are offline.
	Only an explicit enabling of the disabled threads will fix this behavior.
	While we cannot pinpoint it to either Linux OS or AMD processor, we assume that it is the interaction between both, elevating some disabled hardware threads to C1.

	\begin{figure*}[b]
\addtocounter{figure}{1}
		\subfloat[\label{fig:rapl-pkg}Reference vs. RAPL Package]{
			\includegraphics[width=.75\columnwidth]{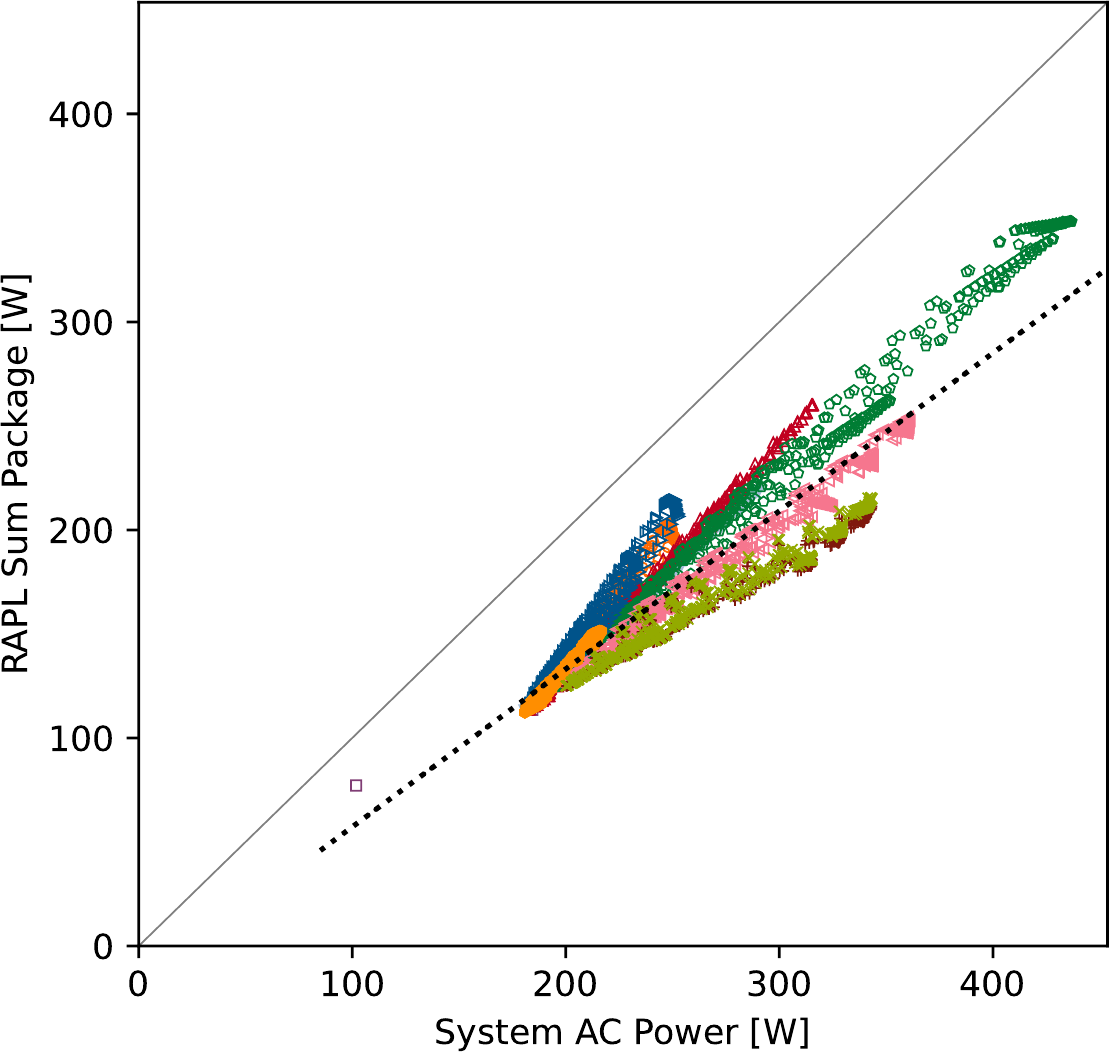}
		}
		\hfill
		\subfloat[\label{fig:rapl-core-pkg}RAPL core vs. RAPL package]{
			\includegraphics[width=.75\columnwidth]{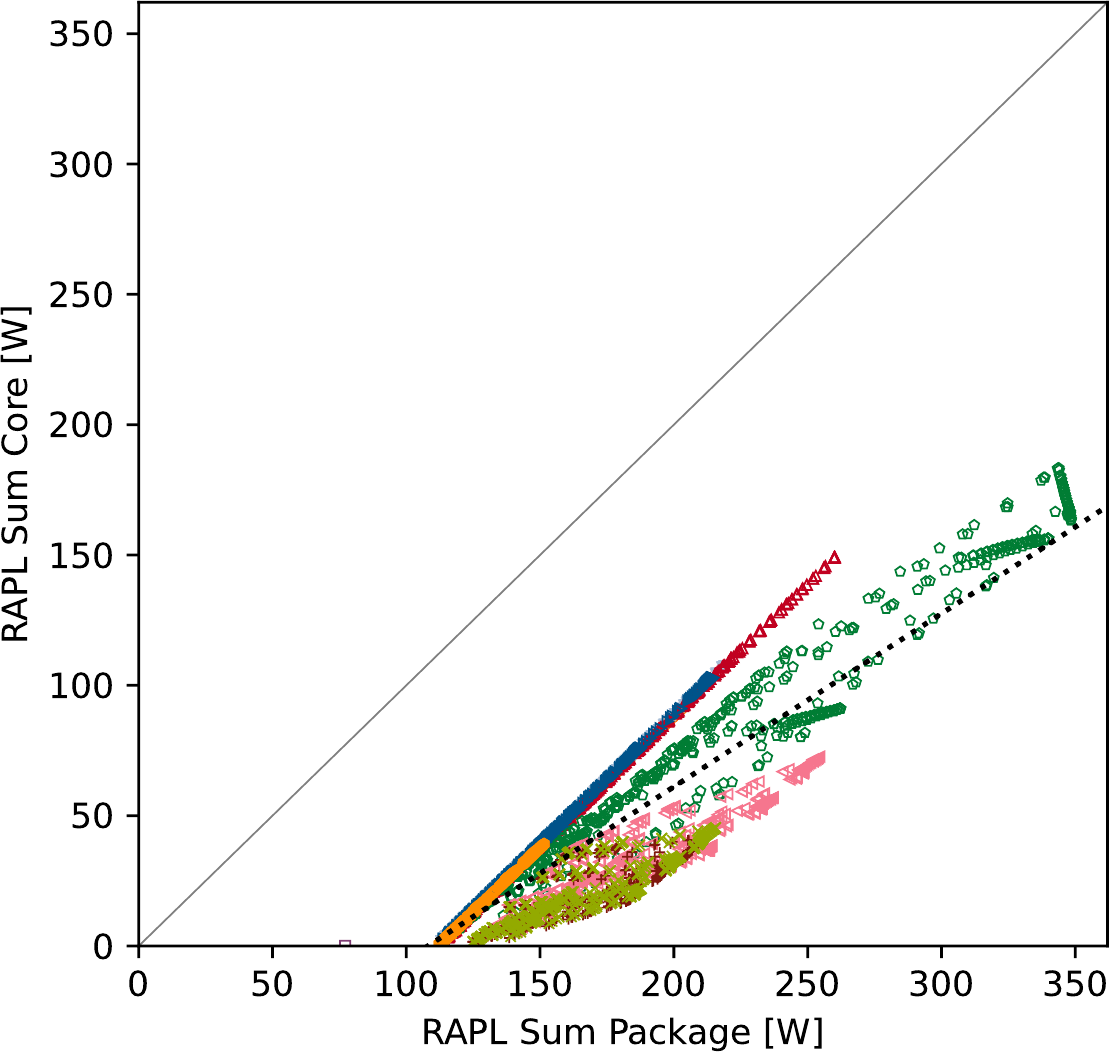}
		}
	    \hfill
		\includegraphics[width=.275\columnwidth,raise=1.1cm]{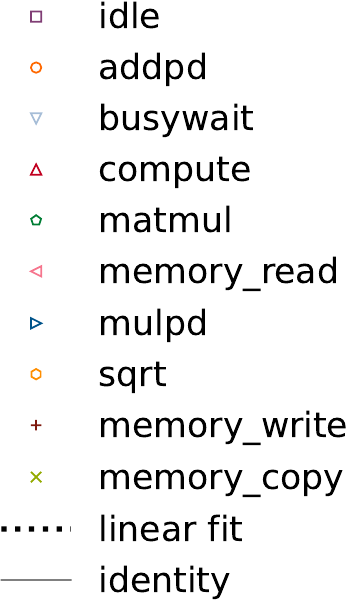}
		\caption{\label{fig:rapl}Readings of RAPL on AMD Epyc 7502 and the AC reference measurements in relation to each other.}
	\end{figure*}
	
		\begin{figure}[t]
\addtocounter{figure}{-2}
		\center
		~
		\hfill
		\subfloat[\label{fig:cstate-c1-local}C1]{
			\includegraphics[height=4.0cm]{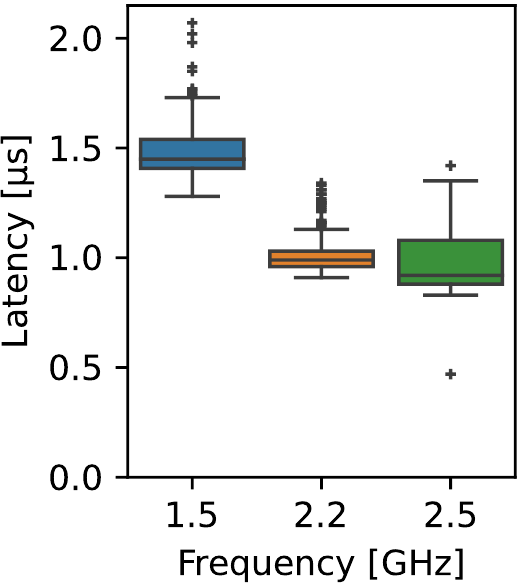}
		}
		\hfill
		\subfloat[\label{fig:cstate-c6-local}C2]{
			\includegraphics[height=4.0cm]{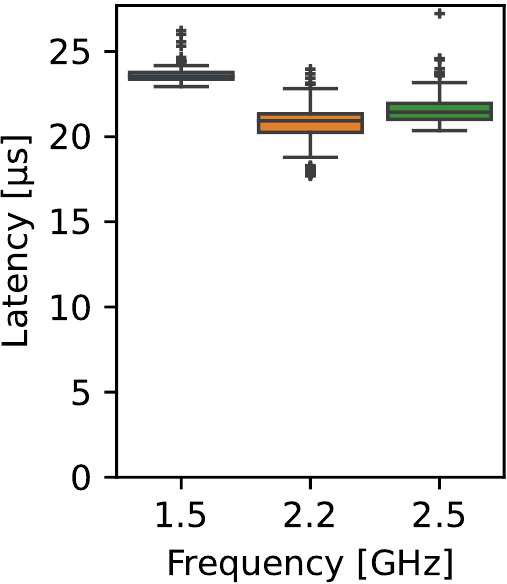}
		}
		\hfill
		~
		\caption{\label{fig:cstate-transition}C-state transitions (caller and callee in same CCX).}
\addtocounter{figure}{1}
	\end{figure}
	
	\subsection{Power State Transition Times}
		
	To determine the transition times for idle states, we use the workload Ilsche et al. explained in \cite{Ilsche_2018_Cstate}.
	However, we had to change the logged event for triggering the transition to \textit{sched\_waking}, since the newer Linux kernel does not report \textit{sched\_wake\_idle\_without\_ipi}, which Ilsche et al. used for the test case.
	The remaining setup stays the same:
	Two threads are started: caller and callee.
	While the callee is idling via \texttt{pthread\_cond\_wait}, the caller wakes it with \texttt{pthread\_cond\_signal}.
	We schedule the threads within a CCX for local measurements and on one core of each of the two sockets for remote measurements.
	We also measure the influence of frequencies and take \num{200} samples for each combination of C-state, frequency, and local/remote-scenario.
	Results for local transition times are depicted in \figref{cstate-transition}.

	The outliers can be attributed to the measurement, which runs on the same resources as the test workload and therefore influences the results.
	Also, the depicted C-state is the one requested by the OS, not necessarily the one that is realized by the hardware.
	The latency for returning from C1 is consistent with the value reported by hardware with $\sim$\SI{1}{\micro\second} at \SI{2.2}{\giga\hertz} and \SI{2.5}{\giga\hertz} and \SI{1.5}{\micro\second} at \SI{1.5}{\giga\hertz}.
	The C2 latency is between \SI{20}{\micro\second} and \SI{25}{\micro\second} and significantly lower than reported to the OS (\SI{400}{\micro\second}).
	However, this value could significantly increase when package C-states are used, which disable additional processor components.
	This case is not measurable with the used methodology since the active caller would prevent package C-states.
	
	Transition times for remote configurations only add a small overhead ($\sim$\SI{1}{\micro\second}) to the results shown in these diagrams and are therefore not presented in this paper.
	However, this validates the finding from \secref{idle-power}: package C-states are not used as long as a single core (which runs the caller thread) in the system is active.

	\section{Integrated Energy Measurement with RAPL}
	\label{sec:rapl}
	In this section, we analyze AMD's RAPL implementation.
	We se\-p\-a\-rate\-ly analyze its quality with a high-level focus on the executed workload and the detailed impact of input data on instruction-level power consumption.
	We measured an update rate of \SI{1}{\milli\second} for RAPL by polling the MSRs via the \texttt{msr} kernel module, which meets the specification for Intel processors.

	\subsection{Quality of the Integrated Power Measurement}
	\label{sec:rapl-roco}

	To analyze the accuracy of RAPL readings, we follow the methodology presented by Hackenberg et al.~\cite{Hackenberg_2015_Haswell}.
	We execute a set of ex\-peri\-ments, where each forms a particular combination of workload, thread placement, frequency, and enabled C-states, for a duration of \SI{10}{\second}.
	We record RAPL package energy measurements, RAPL core information, and system AC power for each workload configuration as described in \secref{test-system}.

	If AMD RAPL would use an accurate measurement that covers all components for which power varies by workload, a single function would map RAPL readings to the reference measurement.
	Instead, \figref{rapl-pkg} is reminiscent of Intel's implementation before Haswell~\cite{Hackenberg_Power_2013}.
	The results indicate that the energy data is modeled, not measured:
	Even workloads that do not use memory (\texttt{sqrt}, \texttt{add\_pd}, \texttt{mul\_pd}) show inconsistent power.
	Further, the energy consumption of memory accesses (e.g., \texttt{memory\_read}, \texttt{memory\_write}) is not fully captured by RAPL. No DRAM domain is available and the RAPL package domain reports significantly lower power compared to the external measurement.
	Considering the different domains, this does not necessarily imply that RAPL readouts are wrong.
	But it shows that they cannot be used to accurately estimate and therefore optimize for total system power, as opposed to Intel systems since Haswell.
	On such systems, this is possible when adding Package and DRAM energy~\cite{Hackenberg_2015_Haswell}.
	
	The comparison in \figref{rapl-core-pkg} reveals that there is a simple relation between the different RAPL measurement domains for compute-only workloads while the power difference for memory-intensive workloads and idle varies.
	This seems intuitive, since a model for package power consumption would include modeled core energy but also shared non-core resources.

		\begin{figure*}[b]
		\centering
		\subfloat[\label{fig:rapl-data-xor-10s-sys}Reference measurement, full system]{
			\includegraphics[width=0.95\columnwidth]{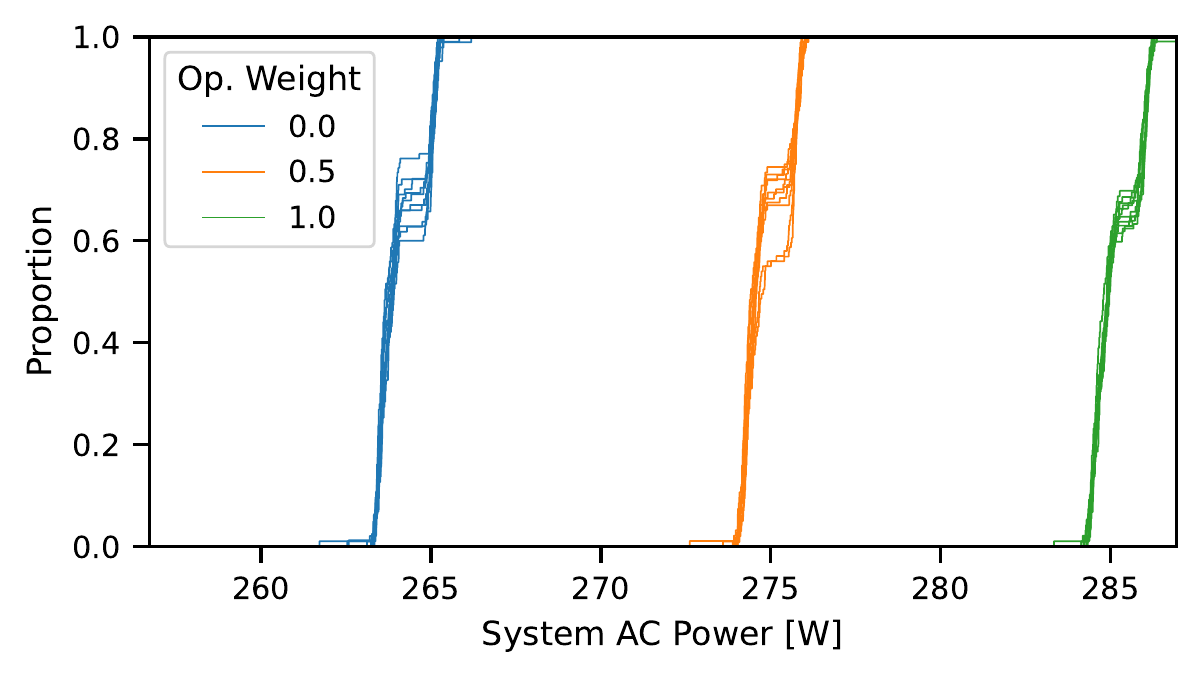}
		}
		\hfill
		\subfloat[\label{fig:rapl-data-xor-10s-core}RAPL, core]{
			\includegraphics[width=0.95\columnwidth]{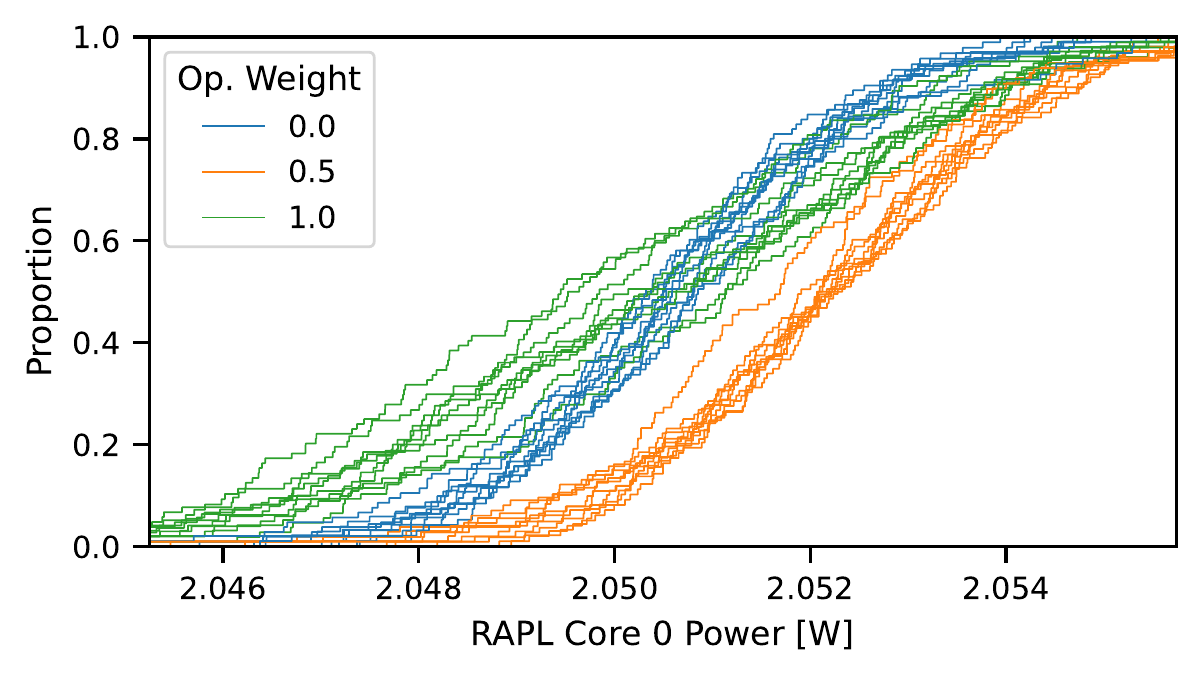}
		}
		\caption{\label{fig:rapl-data}%
			Full system AC and RAPL power consumptions for \texttt{vxorps}.
			Each chart shows the empirical cumulative distribution of ten random sample sets for the three different relative operand Hamming weights.%
		}
	\end{figure*}	
	\subsection{Measurement of Data-dependent Power Consumption}
	\label{sec:rapl-data}
	
	The power consumption for executing a workload does not only depend on the used instructions, but also on the processed data.
	Given data-dependent power differences of up to $\sim$\SI{15}{\percent} full system power~\cite{Schoene_2019_SKL}, data can also have a significant impact on RAPL accuracy.
	Subsequently, applied processor frequencies can be inaccurate, which can lead to an exceeded TDP or performance loss.
	
	Another aspect to consider is the exploitation of RAPL for software-based side-channel attacks as demonstrated on Intel systems.
	In~\cite{Lipp_2020_Platypus}, Lipp et al. also indicate that it could be possible on newer AMD systems.
	On the one hand, the authors distinguish different \emph{instructions} based on RAPL measurements.
	While not always accurate, \figref{rapl} confirms that RAPL on Zen\,2 does reflect the different power consumption of instructions to some extent.
	On the other hand, Lipp et al. use RAPL to distinguish \emph{operands} of instructions.
	To that end, they measure the energy consumption of the \texttt{shr} instruction with RAPL on an AMD Zen\,2 desktop system and show a slightly shifted probability density function for different numbers of set bits (operand Hamming weight).
	
	We measure the instruction power consumption by repeating an unrolled loop of the respective instruction for a fixed number of total instructions.
	Successive instructions use different registers to avoid stalling.
	For each block of instructions, the test application randomly chooses a relative Hamming weight of either \num{0}, \num{0.5}, or \num{1} and executes this configuration on all hardware threads.
	The instruction count is chosen such that each instruction block runs for \SI{10}{\second}.
	Overall, \num{3000} instruction blocks are executed ($\sim$\num{1000} per operand weight).
	The experiment application collects RAPL energy values between instruction blocks.
	Even though the measured duration is very stable, we normalize the energy values to power.
	
	First, we look at a \num{256}-bit \texttt{vxorps} instruction and vary the the operand that determines the toggled bits in the destination register.
	\figref{rapl-data} illustrates the distribution of average power values for the repeated instruction blocks of each operand configuration.
	To avoid smoothing, we use empirical cumulative distribution plots.
	Moreover, to confirm whether the distribution is stable, we separate the samples into ten random subsets and plot the distribution for each subset.
	As can be seen in \figref{rapl-data-xor-10s-sys}, the system power consumption increases with the number of toggled bits with a significant difference of \SI{21}{\watt} (\SI{7.6}{\percent}) with no overlap in distributions.
	The RAPL measurements do not reflect this difference: Their overall averages are within \SI{0.08}{\percent} for different operand weights.
	\figref{rapl-data-xor-10s-core} shows that the distributions are distinguishable but strongly o\-ver\-lapp\-ing.
	Moreover, the clear ordering between operand weights \num{0}, \num{0.5}, \num{1} is not reflected by RAPL.
	
	To contrast the findings of~\cite{Lipp_2020_Platypus}, we also ran the experiment with a \num{64}-bit \texttt{shr} instruction.
	The operand is seeded depending on the selected operand weight and repeatedly shifted by \num{0}.
	The system power consumption averages are much closer within \SI{0.9}{\percent} for different operand weights whereas RAPL core power averages are within \SI{0.015}{\percent} and their distribution overlaps similarly to the previous experiment.
	In all cases, the RAPL package domain measurements behave similarly, albeit with different ordering of distributions.
	
	Primarily, the results show that the RAPL implementation on our system does not correctly represent the impact of data on power consumption, possibly affecting measurement accuracy in workloads with biased data.
	However, it is conceivable that this RAPL implementation could still be used to leak information about the processed data through very small differences in the distribution of power consumption samples.
	The results indicate that this is due to indirect effects, e.g., an increased temperature based on the number of set bits.
	Nevertheless, distinguishing the operand weight from RAPL values on this system would take substantially more samples compared to a physical measurement.
	Moreover, on our test system, RAPL is not accessible to unprivileged users.

	\section{Conclusion and Future Work}
	\label{sec:summary}
	With the Zen\,2 Rome processors, AMD ships a complex x86 architecture, which includes numerous power saving and monitoring mechanisms for an improved energy efficiency.
	In this paper, we provide a detailed analysis of these.
	
	To maximize energy efficiency, users and administrators can take the following measures:
	Hardware threads should not be disabled in the OS as this can disable package C-states and significantly increase idle power consumption under specific circumstances.
	Unused hardware threads should be run at the lowest possible frequency.
	Otherwise, they can raise the frequency of hardware threads on the same core.
	Mixed frequencies within one CCX should be avoided, since this can lead to performance losses on cores with lower frequency settings.
	The processor frequency should be monitored to detect throttling when using \num{256}-bit SIMD instructions.
	This can lead to significant performance degradation, especially for highly parallel HPC codes.
	Energy measurements of AMDs RAPL implementation should be considered inaccurate.
	No DRAM domain is provided, and DRAM energy consumption is not (fully) included in the package domain.
	Therefore, AMD's RAPL is unsuitable to optimize total energy consumption.
	The modeled approach also fails on reflecting the influence of operands, which can also be seen as a benefit when it comes to power measurement based side-channel attacks.
	
	Our findings are valuable for a wide audience:
	Performance models can be improved for a better accuracy, tuning mechanisms can be refined to become more efficient, and operating systems can be optimized to fix or prevent some of the identified peculiarities.
	
	As future work, we will analyze the frequency throttling on processors with more cores.
	We expect a more severe impact, since the ratio of compute to I/O resources is higher.
	We will also analyze the memory architecture and the influence of power saving mech\-a\-nisms on these in higher detail.
	Finally, we plan to analyze why offline hardware threads can prevent the usage of package C-states.

	\section*{Acknowledgments and Reproducibility}
	This work is supported in part by the German Research Foundation (DFG) within the CRC 912 - HAEC.
	
	Measurement programs, raw data, and chart notebooks are available at \url{https://github.com/tud-zih-energy/2021-rome-ee}.
	
	\bibliographystyle{IEEEtran}
	
	\bibliography{paper}
	
\end{document}